\documentclass{article}
\usepackage{PRIMEarxiv}

\usepackage[utf8]{inputenc} 
\usepackage[T1]{fontenc}    
\usepackage{hyperref}       
\usepackage{url}            
\usepackage{booktabs}       
\usepackage{amsfonts}       
\usepackage{nicefrac}       
\usepackage{microtype}      
\usepackage{lipsum}
\usepackage{fancyhdr}       
\usepackage{graphicx}       
\usepackage{booktabs}  
\usepackage{multirow}  
\usepackage{graphicx}  
\usepackage{multirow}  
\usepackage{pifont}    
\usepackage{xcolor}    
\usepackage{colortbl}  
\usepackage{makecell}  

\newcommand{\cmark}{\textcolor{green!60!black}{\ding{51}}} 
\newcommand{\xmark}{\textcolor{red}{\ding{55}}}            

\graphicspath{{media/}}     
\usepackage{amsmath}
\title{Adaptive Confidence Gating in Multi-Agent Collaboration for Efficient and Optimized Code Generation
}

\author{
  Haoji Zhang \\
  University of Electronic Science \\and Technology of China \\
  \texttt{hzhang070@e.ntu.edu.sg} \\
   \And
  Yuzhe Li \\
  University of Electronic Science \\and Technology of China \\
  \texttt{liodysseyz@gmail.com} \\
  \And
  Zhenqiang Liu \\
  University of Electronic Science \\and Technology of China \\
  \texttt{lzq021120@gmail.com} \\
  \And
  Chenyang Liu \\
  University of Electronic Science \\and Technology of China \\
  \texttt{2437345747@qq.com} \\
  \And
  Shenyang Zhang \\
  University of Electronic Science \\and Technology of China \\
  \texttt{zsy0212142022@163.com} \\
  \And
  Yi Zhou\textsuperscript{*} \\
  University of Electronic Science \\and Technology of China \\
  \texttt{zhou.yi@uestc.edu.cn} \\
}

\begin{document}
\maketitle

\renewcommand{\thefootnote}{\fnsymbol{footnote}} 
\footnotetext[1]{Corresponding author} 
\renewcommand{\thefootnote}{\arabic{footnote}} 

\begin{abstract}
While Large Language Models (LLMs) have catalyzed breakthroughs in automated code generation, Small Language Models (SLMs) often encounter reasoning bottlenecks and failure loops when addressing complex logical requirements. To overcome these challenges, this study proposes DebateCoder, a multi-agent collaborative framework specifically engineered to optimize the reasoning potential of SLMs, such as Pangu-1B, within resource-constrained environments. The methodology implements a structured role-playing protocol—comprising User ($A_{UA}$), Technical ($A_{TA}$), and Quality Assurance ($A_{QA}$) agents—coupled with an Adaptive Confidence Gating mechanism using a 95\% threshold to balance accuracy and inference efficiency. Additionally, a multi-turn deliberation module and a reviewer-guided analytical debugging loop are introduced to facilitate orthogonal pre-generation debate and post-generation refinement. Quantitative evaluations on HumanEval and MBPP benchmarks reveal that DebateCoder achieves a Pass@1 accuracy of 70.12\% on HumanEval, significantly surpassing the MapCoder framework while reducing API overhead by approximately 35\%. These findings demonstrate that sophisticated collaborative protocols can effectively mitigate the inherent limitations of small-parameter models, offering a scalable and efficient alternative for high-quality automated software engineering.
\end{abstract}

\keywords{Multi-Agent Collaboration \and Code Generation \and Small Language Model}

\section{Introduction}
In recent years, the field of automated code generation has witnessed significant breakthroughs driven by Large Language Models (LLMs)~\cite{Jiang2024A,Huynh2025Large,Dong2023Self-Collaboration, Nadǎş2025Synthetic}. To tackle the challenges of complex logical reasoning and self-correction in programming tasks, researchers have increasingly shifted their focus from single-model inference to Multi-Agent Collaboration frameworks. State-of-the-art frameworks, such as ChatDev~\cite{qian2023chatdev}, MetaGPT~\cite{hong2024metagpt}, and MapCoder~\cite{mapcoder_citation}, enhance the quality and reliability of generated code by simulating the specialized roles found in human software development lifecycles (e.g., product managers, programmers, and testers).

While multi-agent systems demonstrate superior accuracy, existing high-performance frameworks are predominantly dependent on massive models with hundreds of billions of parameters (e.g., GPT-4, Claude 3.5). In the domain of multi-agent code generation, the utilization of Small Language Models (SLMs) for collaborative tasks remains remarkably scarce and presents significant challenges. This scarcity stems from two primary hurdles: first, SLMs often encounter "reasoning bottlenecks" when handling multi-turn dialogues and complex logic, making it difficult to maintain contextual consistency~\cite{He2024LLM-Based}; second, SLMs frequently fall into "failure loops" during tool invocation and self-debugging phases—namely, the inability to accurately identify and rectify their own logical errors.~\cite{Ashrafi2025Enhancing, Xiong2025ResilioMate:, Chen2025A} Given the constraints of computational resources and inference costs, achieving efficient and precise agent collaboration within a limited parameter scale has become a critical problem to be solved.

To address these challenges, we propose \textbf{DebateCoder}, an efficient collaborative framework specifically designed for small-parameter models. The core motivation of this work is to maximize the reasoning potential of SLMs by optimizing inter-agent interaction strategies without increasing model size. We utilize \textbf{Pangu-1B}~\cite{chen2025pangu} as the backbone model and introduce a suite of innovative mechanisms to overcome the inherent limitations of small models in complex code generation.

The primary contributions of this paper are as follows:
\begin{itemize}
    \item \textbf{DebateCoder Framework:} We propose DebateCoder, a specialized framework comprising a User Agent ($A_{UA}$), Technical Agent ($A_{TA}$), and Quality Assurance Agent ($A_{QA}$). By implementing a structured role-playing and competitive debate protocol, the framework successfully breaks through the reasoning bottlenecks typically encountered by small models in complex programming logic.
    \item \textbf{Adaptive Confidence Gating and Synergistic Debugging:} We introduce an Adaptive Confidence Gating mechanism with a 95\% threshold to balance accuracy and efficiency, reducing API overhead by approximately 35\%. Furthermore, we design a multi-turn deliberation module coupled with a Code Reviewer-driven debugging agent, where pre-generation debate and post-generation refinement work in an orthogonal and complementary manner to ensure code robustness.
    \item \textbf{Extensive Performance Validation:} Across standard benchmarks such as HumanEval and MBPP, DebateCoder consistently outperforms direct generation methods and the traditional MapCoder framework. This demonstrates that small models, when guided by well-designed collaborative protocols, are capable of completing high-quality code generation tasks.
\end{itemize}
\section{Related Work}
\label{sec:related_work}

Current research in automated software engineering is characterized by a rapid transition from single-model synthesis to agentic workflows and collaborative reasoning. Our work is situated at the convergence of three dynamic fields: neural code generation with self-correction, multi-agent collaboration (MAC), and reasoning optimization for Small Language Models (SLMs).

\subsection{Code Generation and Self-Correction Mechanisms}
The advent of transformer-based models has revolutionized code generation. Early foundational models such as Codex~\cite{chen2021codex} and StarCoder~\cite{li2023starcoder} demonstrated the efficacy of training on massive code corpora. More recently, open-source models like WizardCoder~\cite{luo2023wizardcoder}, DeepSeek-Coder~\cite{guo2024deepseek}, and StarCoder2~\cite{lozhkov2024starcoder2} have significantly narrowed the gap with closed-source counterparts. 

To address complex logic, researchers have leveraged In-Context Learning (ICL)~\cite{brown2020language} and advanced prompting strategies. Chain-of-Thought (CoT)~\cite{wei2022chain} and its zero-shot variant~\cite{kojima2022large} encourage models to decompose problems into intermediate steps. However, single-pass generation often proves insufficient for rigorous engineering tasks. This has led to the development of self-correction mechanisms. Techniques like Self-Repair~\cite{olausson2024demystifying} and Self-Refine~\cite{madaan2024self} prompt models to critique their outputs. Reflexion~\cite{shinn2023reflexion} further advances this by using verbal reinforcement learning to maintain a memory of past mistakes. Despite these innovations, studies indicate that SLMs often struggle with intrinsic self-correction due to limited reasoning depth, leading to ``failure loops''~\cite{zhang2023selfedit}. Our work builds on these foundations but introduces an adversarial debate mechanism to drive correction externally.

\subsection{Multi-Agent Collaboration (MAC) Frameworks}
To transcend individual model limitations, the field has shifted toward Multi-Agent Collaboration (MAC). Inspired by human software development, frameworks like ChatDev~\cite{qian2023chatdev} and MetaGPT~\cite{hong2024metagpt} assign specialized roles (e.g., architect, engineer, tester) to agents, enforcing Standardized Operating Procedures (SOPs).

Recent advancements have focused on specialized workflows. OpenHands (formerly OpenDevin)~\cite{wang2025openhands} provides a generalist platform for agents to interact with command lines and browsers. In competitive programming, MapCoder~\cite{islam2024mapcoder} employs a master-worker architecture to plan and debug solutions iteratively. For maintenance, AutoCodeRover~\cite{zhang2024autocoderover} leverages abstract syntax tree (AST) analysis to localize faults. While these frameworks demonstrate superior accuracy, they predominantly rely on massive, closed-source models (e.g., GPT-4). Our approach democratizes these capabilities by optimizing MAC protocols specifically for SLMs, utilizing structure-aware debate to compensate for lower parameter counts.

\subsection{Debate, Reasoning, and SLM Optimization}
Enhancing the reasoning capabilities of smaller models is critical for efficient deployment. Multi-agent debate (MAD) has emerged as a powerful technique to improve factuality and logic. Du et al.~\cite{du2023improving} showed that divergent views among agents can converge on correct answers, mitigating hallucinations. This is further supported by Liang et al.~\cite{liang2023encouraging}, who demonstrated that divergent thinking promotes comprehensive problem exploration.

Recent works have tailored these mechanisms specifically for complex reasoning. For instance, DMAD~\cite{liu2025dmad} fosters diversity by assigning distinct reasoning strategies (e.g., CoT vs. Program-of-Thoughts) to different agents. Parallel efforts in SLM optimization, such as Skip-Thinking~\cite{chen2025skip}, employ chunk-wise distillation to transfer complex reasoning patterns from teacher models to SLMs. However, applying debate mechanisms directly to SLMs presents challenges. Unlike giant models, SLMs require more structured guidance to avoid ``debate collapse,'' where agents reinforce each other's errors rather than correcting them. To address this, we introduce an \textit{Adaptive Confidence Gating} mechanism and a structured \textit{Reviewer-Debugger} protocol.





\section{Methods}

In this section, we first delineate the specialized agent roles and their respective prompting architectures within the proposed framework in Section 3.1. Subsequently, we detail the operational pipeline of DebateCoder, including the consensus-driven initial planning and confidence-based gating mechanism in Section 3.2, the iterative cross-agent debate protocol in Section 3.3, the synthesized fusion and code generation process in Section 3.4, and the reviewer-guided analytical debugging loop tailored for programming challenges in Section 3.5.

\subsection{DebateCoder Strategy}
Our objective is to develop a multi-agent code generation framework, \textbf{DebateCoder}, specifically engineered to handle the intricate logic required for programming. To achieve this, our framework replicates the collaborative environment of a professional software development ecosystem through a set of specialized agents $\mathcal{A} = \{A_{UA}, A_{TA}, A_{QA}\}$. We devise a structured pipeline that cascades these agents to foster a multi-dimensional consensus on problem requirements and algorithmic design. The framework orchestrates the synthesis process by simulating three core expert perspectives:

\begin{itemize}
    \item \textbf{User Agent ($A_{UA}$):} Acts as a product owner focusing on \textit{Functionality Completeness and Usability}. Its primary responsibility is to ensures the proposed solution fulfills all user requirements while maintaining an intuitive and clear interface alignment.
    \item \textbf{Technical Agent ($A_{TA}$):} Operates as a lead architect prioritizing \textit{Technical Feasibility and Performance Efficiency}. It ensures the implementation is algorithmically sound, leveraging optimal data structures and design patterns to achieve superior time and space complexity.
    \item \textbf{QA Agent ($A_{QA}$):} Functions as a quality assurance engineer emphasizing \textit{Robustness and Reliability}. It scrutinizes the plan for potential vulnerabilities, ensuring thorough input validation and comprehensive exception handling for boundary conditions and edge cases.
\end{itemize}

As illustrated in Figure, the execution of \textbf{DebateCoder} is organized into a synergistic four-stage pipeline: (1) \textbf{Parallel Initialization and Confidence Evaluation}, (2) \textbf{Iterative Multi-turn Deliberation}, (3) \textbf{Plan Fusion and Code Generation}, and (4) \textbf{Reviewer-Guided Debugging}. This structured flow ensures that the final code is not only functionally correct but also resilient and technically optimized.

\begin{figure}[htbp]
  \centering
  \includegraphics[width=0.8\textwidth]{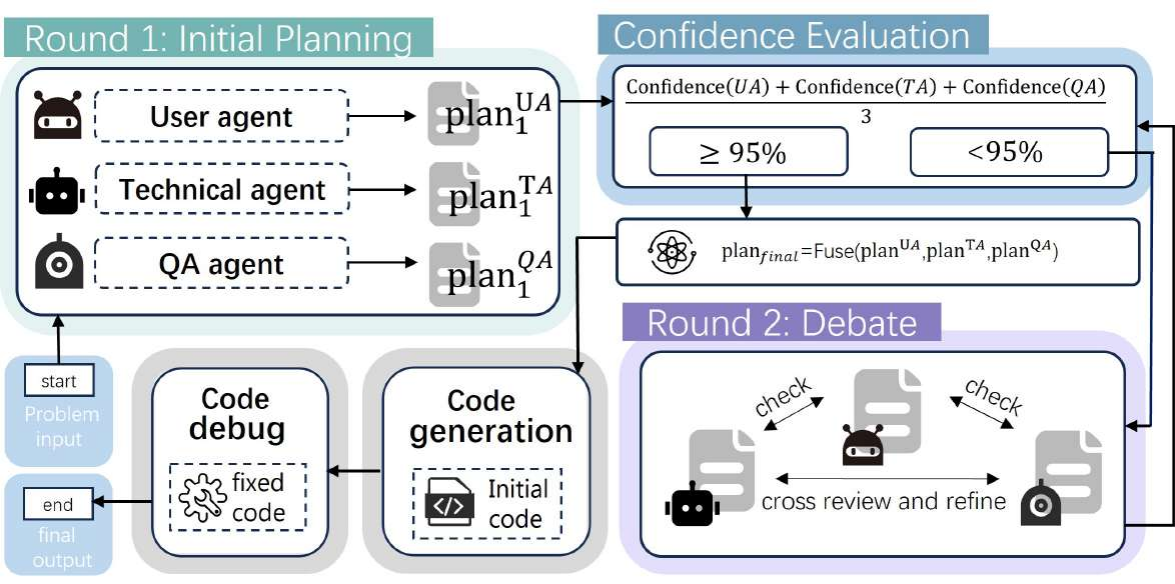} 
  \caption{overview of debate coder}
  \label{fig:my_diagram}
\end{figure}

\subsection{Initial Planning and Adaptive Confidence Gating}
\begin{figure}[h]
    \centering
    \includegraphics[width=0.5\linewidth]{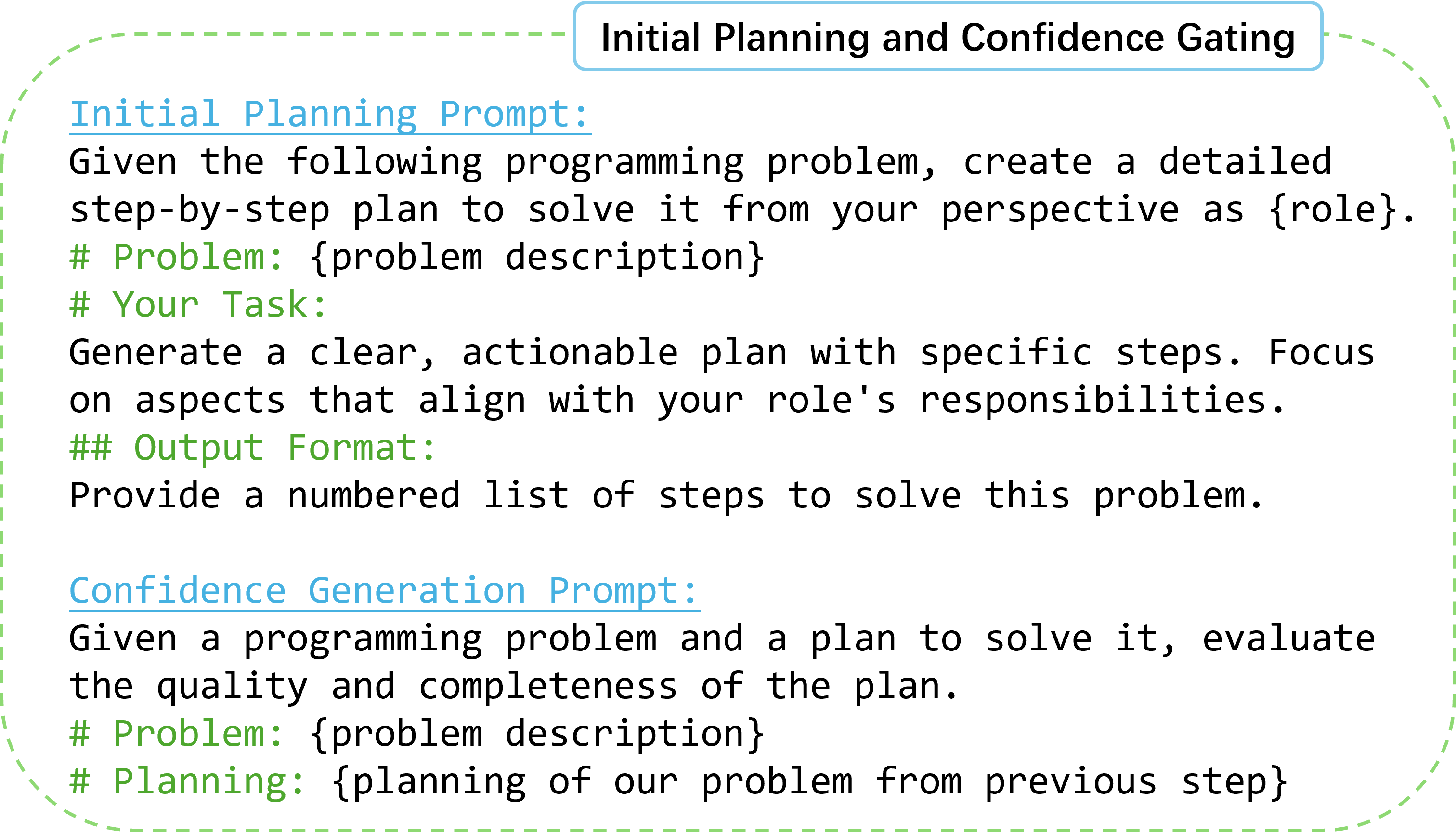}
    \caption{Prompt for Initial Planning and Confidence Gating.}
    \label{fig:fig3}
\end{figure}

The first phase of the pipeline, Parallel Initialization (Round $R_1$), involves each agent $A_i \in \mathcal{A}$ independently generating a candidate plan $L_{i,1}$ tailored to its specific persona. Unlike monolithic prompting, this stage encourages divergent thinking by allowing agents to prioritize their respective domain-specific metrics. To facilitate downstream processing, our designed prompt in Figure \ref{fig:fig3} explicitly instructs the LLM to generate both a detailed plan and a utility-based confidence score $c_{i,1} \in [0, 100]$, which reflects the perceived solvability and clarity of the task $P$. Formally:
\begin{equation}
    (L_{i,1}, c_{i,1}) = \text{LLM}(P, \text{Persona}_i)
\end{equation}

To balance the trade-off between synthesis quality and computational overhead, we implement a \textbf{Confidence-based Gating} mechanism. We calculate the collective confidence $\Gamma = \frac{1}{|\mathcal{A}|} \sum c_{i,1}$ to serve as a heuristic for problem difficulty. If $\Gamma$ exceeds a predefined threshold $\tau$ (e.g., $\tau = 95\%$), the system classifies the problem as "low-complexity," assuming the initial independent plans have already reached a high-quality consensus. In such scenarios, the framework bypasses all subsequent debate rounds and proceeds directly to the fusion stage. This early termination strategy significantly reduces inference latency and token consumption for straightforward tasks while preserving the multi-agent deliberation capacity for truly challenging instances.

\subsection{Iterative Cross-Agent Deliberation}
For problems where the collective confidence fails to meet the threshold ($\Gamma < \tau$), the framework initiates an iterative debate phase for a maximum of $R$ rounds (typically $R=3$). In each round $r \in [2, R]$, every agent $A_i$ is presented with the complete set of plans generated by its peers in the previous round, $\mathcal{L}_{r-1} = \{L_{j,r-1} \mid j \in \mathcal{A}\}$. This peer-review process compels each agent to perform a comparative analysis, identifying architectural weaknesses in its own prior logic and assimilating superior strategies from other roles. The refinement process is defined as:
\begin{equation}
    L_{i,r} = \text{LLM}(P, \mathcal{L}_{r-1}, \text{Persona}_i)
\end{equation}

This phase facilitates a dynamic convergence of specialized knowledge. For example, the Technical Agent ($A_{TA}$) may revise its choice of data structure to accommodate a specific boundary constraint highlighted by the QA Agent ($A_{QA}$). Simultaneously, the User Agent ($A_{UA}$) ensures that any technical optimizations do not deviate from the original functional intent of the user. After each round, the confidence scores are re-evaluated, allowing the system to exit the debate early if a consensus is reached before the maximum round $R$ is exhausted. This collaborative reasoning ensures a comprehensive solution that mitigates the individual "blind spots" often found in single-agent Chain-of-Thought paths.

\subsection{Consensus Synthesis and Code Generation}
The final planning stage involves a \textbf{Consensus Synthesis} process. A specialized Synthesis Agent $\mathcal{S}$ is tasked with aggregating the terminal round of plans $\mathcal{L}_R$ into a single, high-fidelity Master Plan $P^*$. Formally:
\begin{equation}
    P^* = \text{LLM}(\text{Synthesize}(\mathcal{L}_R))
\end{equation}

The synthesizer is instructed to resolve any lingering logical contradictions by prioritizing the most robust, efficient, and functionally complete components identified during the debate. Rather than a simple text concatenation, this synthesis represents a logical integration that balances the divergent priorities of the UA, TA, and QA agents into a self-consistent roadmap. Once $P^*$ is established, it serves as the definitive instruction set for the Coding Agent to produce the executable code. By decomposing the planning process into specialized deliberation followed by synthesized fusion, \textbf{DebateCoder} effectively minimizes the "Failure Loops" encountered in complex  programming, ensuring that the generated implementation is functionally aligned, technically optimized, and resilient to edge cases.

\subsection{Reviewer-Guided Debugging}
In the MapCoder framework~\cite{mapcoder_citation}, the Debugging Agent primarily relies on problem descriptions and test reports to modify code. However, for smaller models like Pangu-1B, simply using binary signals such as "pass/fail" or lengthy error stacks often makes it difficult to pinpoint complex logical bugs and may even introduce new errors through blind modifications (Failure Loop). To address this issue, we introduced a Code Reviewer role with analytical capabilities during the debugging phase.

\begin{figure}
    \centering
    \includegraphics[width=0.5\linewidth]{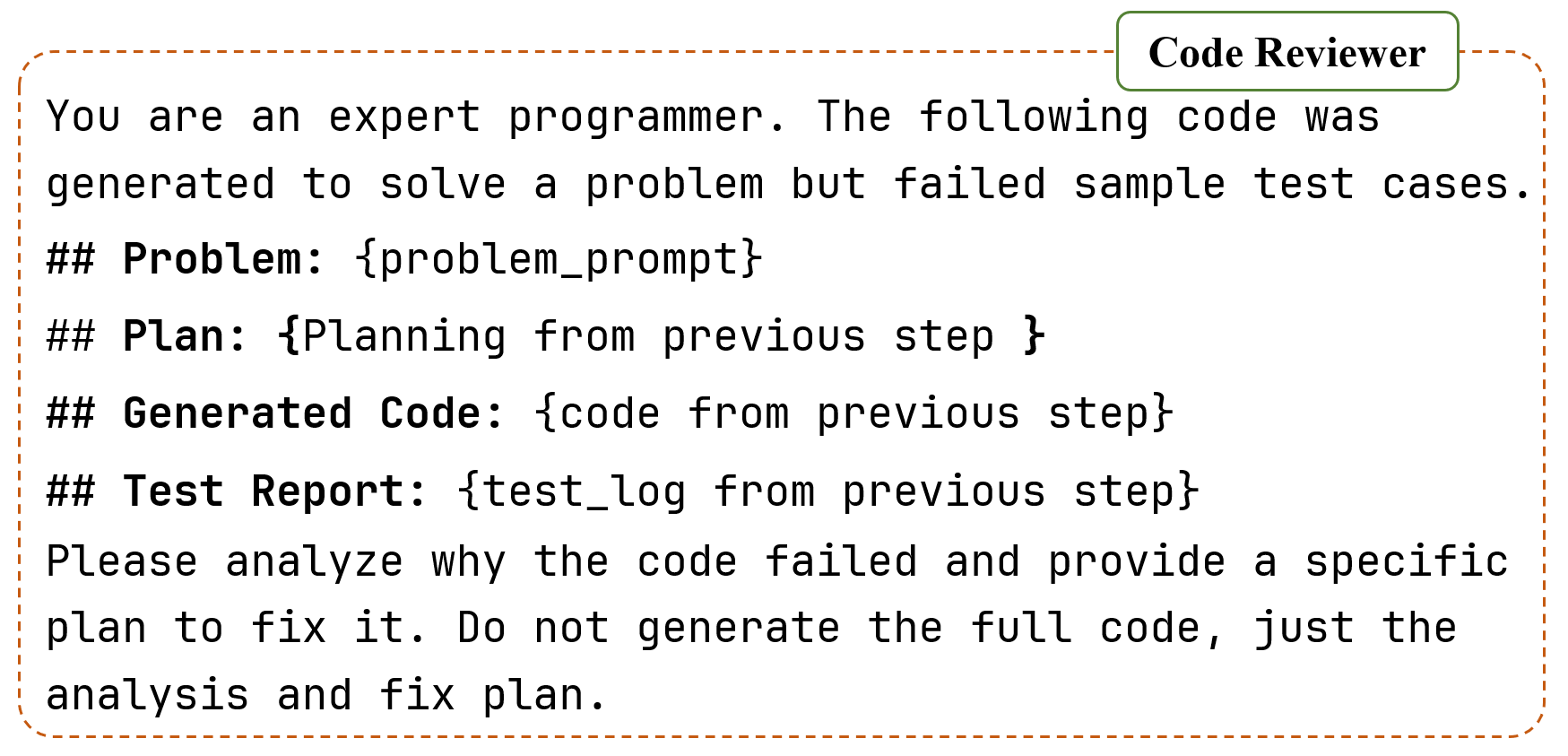}
    \caption{Prompt for Code Reviewer.}
    \label{fig:CodeReviewer}
\end{figure}

\begin{figure}
    \centering
    \includegraphics[width=0.5\linewidth]{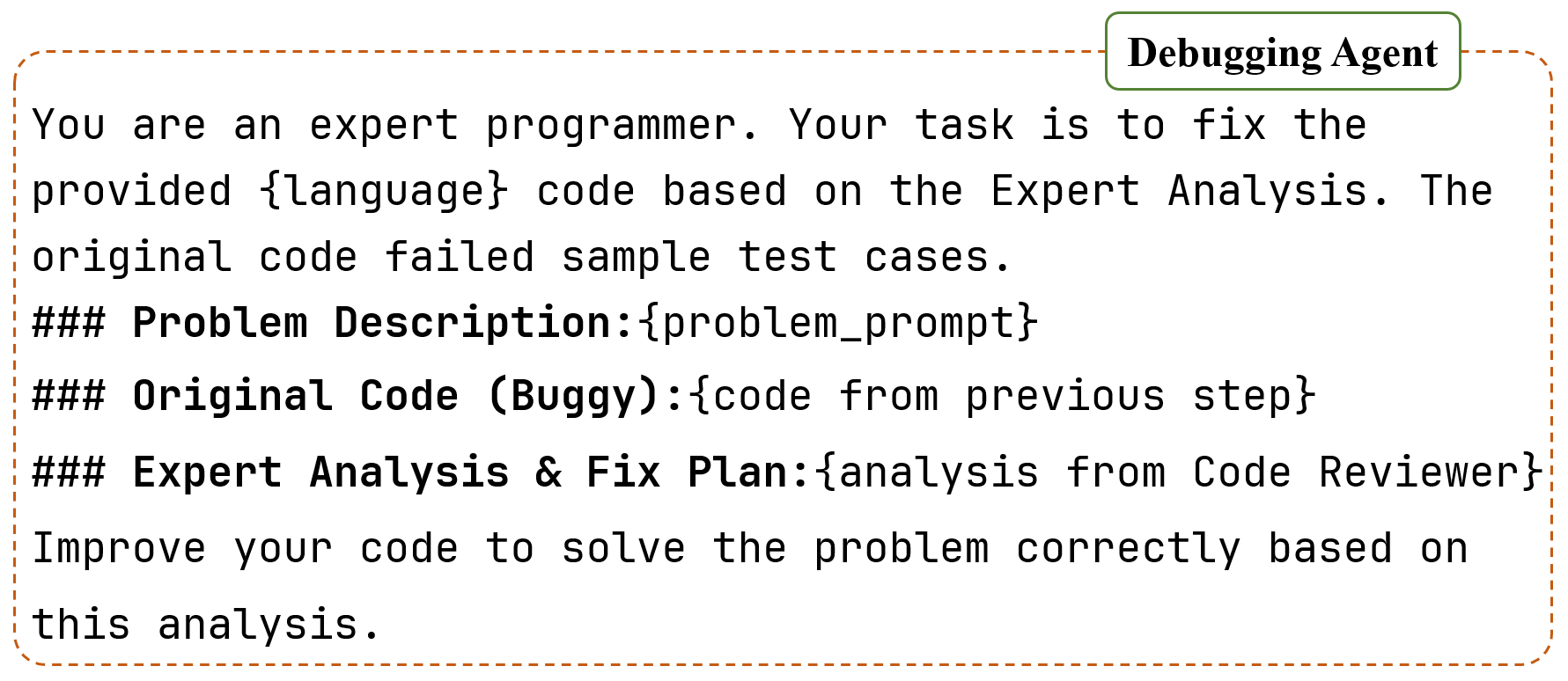}
    \caption{Prompt for Debugging Agent.}
    \label{fig:DebuggingAgent}
\end{figure}

The core responsibility of the Code Reviewer is not to directly rewrite the code but to simulate the behavior of human engineers during the code review process: cause analysis and fix plan formulation. Specifically, when a test fails, the reviewer receives a triplet consisting of \textbf{the problem prompt, code, and test log}. The generated analysis and fix plan are then used to assist the Debugging Agent in making targeted corrections. The prompts for Code Reviewer and Debugging Agent are illustrated in Figure \ref{fig:CodeReviewer}  and Figure \ref{fig:DebuggingAgent}


\section{Experimental Settings}
\textbf{Models.} In our experiments, we use \textbf{Pangu-1B} as the base model for all the agents, deployed on bare-metal servers equipped with \textbf{8 Ascend 910B AI accelerators}.

\textbf{Datasets.} We evaluate the performance of our proposed method on four code generation benchmarks: \textbf{HumanEval}\cite{chen2021evaluatinglargelanguagemodels}, \textbf{MBPP}\cite{austin2021programsynthesislargelanguage}, and their extended variants, \textbf{HumanEval ET}\cite{dong2024codescoreevaluatingcodegeneration} and \textbf{MBPP ET}\cite{dong2024codescoreevaluatingcodegeneration}. These datasets provide a comprehensive assessment of the model's coding capabilities across varying difficulty levels and test case scenarios. 

\textbf{Baselines.} To validate the effectiveness of our approach, we compare it against two primary baselines:
\textbf{Direct:} A standard zero-shot setting where the base model (Pangu-1B) directly generates code based on the problem description without any additional guidance or intermediate steps.
\textbf{MapCoder} \cite{mapcoder_citation}: A competitive method that utilizes multi-agent collaboration to improve code generation quality.

\section{Experimental Results}

\subsection{Main Results }
We compare  DebateCoder against two baselines: \textbf{Direct}, \textbf{MapCoder}, across four benchmarks on the Pangu-1B backbone. The main results are reported in \textbf{Table \ref{tab:main_results}}. 

\textbf{Overall, DebateCoder achieves the best average performance (58.91\%)}, surpassing MapCoder (55.95\%) and significantly outperforming the Direct approach (39.51\%). This indicates that the debate mechanism enhances the model's code generation capabilities effectively. 

Specifically, on the \textbf{HumanEval and HumanEval ET benchmarks}, DebateCoder demonstrates superior performance. On HumanEval, our method achieves a Pass@1 accuracy of \textbf{70.12\%}, outperforming MapCoder by a substantial margin of \textbf{8.53\%}. Similarly, on HumanEval ET, DebateCoder leads with \textbf{60.98\%}, validating its robustness in handling complex logic generation tasks. 

Regarding the \textbf{MBPP and MBPP ET datasets}, DebateCoder shows \textbf{competitive performance}. While it slightly trails MapCoder, it achieves a remarkable improvement over the Direct baseline (e.g., \textbf{63.22\% vs. 28.46\%} on MBPP). This suggests that while MapCoder is highly optimized for specific patterns in MBPP, DebateCoder maintains a high level of accuracy and offers a more balanced performance across diverse benchmarks. 

\begin{table*}[t]
\centering
\caption{Performance of Pangu-1B on Code Generation Benchmarks using Various Approaches. }
\label{tab:main_results}
\resizebox{\textwidth}{!}{%
\begin{tabular}{llccccc}
\toprule
\multirow{2}{*}{\textbf{LLM}} & \multirow{2}{*}{\textbf{Approach}} & \multicolumn{5}{c}{\textbf{Dataset}} \\
\cmidrule(lr){3-7}
 & & \textbf{HumanEval} & \textbf{HumanEval ET} & \textbf{MBPP} & \textbf{MBPP ET} & \textbf{AVG} \\
\midrule
\multirow{3}{*}{Pangu-1B} 
 & Direct      & 59.15\%& 53.04\% & 28.46\% & 17.38\% & 39.51\%\\
 & MapCoder    & 61.59\% & 53.66\% & 65.49\% & 43.07\% & 55.95\% \\
 & DebateCoder & \textbf{70.12\%} & \textbf{60.98\%} & 63.22\% & 41.31\% & \textbf{58.91\%} \\
\bottomrule
\end{tabular}%
}
\end{table*}

\subsection{Efficiency Analysis }
We detail in Table \ref{tab:token_cost} the specifics of computational resource consumption of Pangu-1B during the execution of evaluation tasks with different methods, mainly focus on the number of API calls and token usage (prompts and completions). 

\textbf{In terms of API efficiency, DebateCoder demonstrates a clear advantage over MapCoder.} On the HumanEval benchmark, our method reduces the average number of API calls from 12.45 (MapCoder) to 8.12, representing a reduction of approximately \textbf{35\%}. A similar trend is observed on the MBPP dataset (9.37 vs. 12.13). This reduction is significant because fewer API calls imply lower network latency and reduced interaction overhead, making the generation process more streamlined.

\textbf{Regarding token consumption}, the Completion Tokens remain comparable between DebateCoder and MapCoder (e.g., 12.4k vs. 11.2k on HumanEval). This suggests that our debate mechanism improves code quality without generating unnecessarily verbose outputs.

\textbf{It is worth noting that DebateCoder utilizes more Prompt Tokens} (29.7k on HumanEval) compared to other baselines. This increase is expected and justifiable: the debate mechanism relies on preserving the history of multi-agent interactions to facilitate effective reasoning and self-correction. We argue that trading computational context (Prompt Tokens) for reduced interaction latency (API Calls) and superior accuracy (as shown in Table 1) is a favorable trade-off in modern LLM applications.

\begin{table*}[t]
\centering
\caption{Comparison of API calls and token consumption across different methods using Pangu-1B. The units for tokens are in thousands (k).}
\label{tab:token_cost}
\resizebox{\textwidth}{!}{%
\begin{tabular}{llcccccc}
\toprule
\multirow{3}{*}{\textbf{LLM}} & \multirow{3}{*}{\textbf{Approach}} & \multicolumn{6}{c}{\textbf{Dataset}} \\
\cmidrule(lr){3-8}
 & & \multicolumn{3}{c}{\textbf{HumanEval}} & \multicolumn{3}{c}{\textbf{MBPP}} \\
\cmidrule(lr){3-5} \cmidrule(lr){6-8}
 & & API Calls & Prompt Token (k)& Compl.  Token(k)& API Calls & Prompt Token (k)& Compl. Token(k)\\
\midrule
\multirow{3}{*}{Pangu-1B}& Direct      & 1     & 0.14 & 1.9  & 1     & 0.07 & 0.89 \\
 & MapCoder    & 12.45 & 9.4  & 11.2 & 12.13 & 8.89 & 10.4 \\
 & DebateCoder & 8.12& 29.7 & 12.4 & 9.37& 26.3 & 11.4 \\
\bottomrule
\end{tabular}%
}
\end{table*}

\subsection{Ablation Study }

To verify the effectiveness of each component in \textbf{DebateCoder}, we conducted an ablation study on the HumanEval dataset. The results are summarized in Table \ref{tab:ablation}. We use the Direct generation method as our baseline, which achieves a Pass@1 accuracy of 59.15\%. 
The \textbf{Plan Module} (MultiAgentDebate) improves the baseline accuracy by \textbf{4.87\%} (from 59.15\% to 64.02\%), confirming the value of the debate mechanism. The \textbf{Debug Module} shows even stronger performance, especially when the Code Reviewer and Debugging Agent are combined (67.07\%), underscoring the necessity of structured code review. The best result is achieved by the full \textbf{DebateCoder} framework (\textbf{70.12\%}), demonstrating that the pre-generation debate and post-generation debugging are orthogonal and complementary strategies for maximizing code generation accuracy. 

\begin{table}[t]
\centering
\caption{Ablation study of different modules on the HumanEval dataset. }
\label{tab:ablation}
\resizebox{0.6\columnwidth}{!}{
\begin{tabular}{c|c|c|c} 
\toprule
\rowcolor{gray!15} 
\textbf{Plan Module} & \multicolumn{2}{c|}{\textbf{Debug Module}} & \\
\rowcolor{gray!15} 
MultiAgentDebate & Code Reviewer & Debugging Agent & \multirow{-2}{*}{\textbf{Pass@1}} \\ 
\midrule
\xmark & \xmark & \xmark & 59.15\% \\
\cmark & \xmark & \xmark & 64.02\% \\
\xmark & \xmark & \cmark & 61.59\% \\
\xmark & \cmark & \cmark & 67.07\% \\
\rowcolor{green!5} 
\cmark & \cmark & \cmark & \textbf{70.12\%} \\
\bottomrule
\end{tabular}%
}
\end{table}

\section{Conclusion}
In this paper, we presented DebateCoder, an efficient multi-agent collaborative code generation framework specifically designed for Small Language Models (SLMs). To address the reasoning bottlenecks and failure loops inherent in SLMs during complex programming tasks, we leveraged Pangu-1B as the backbone model and introduced several innovative mechanisms. Through the Adaptive Confidence Gating mechanism (with a 95\% threshold), the framework achieves a critical balance between generation efficiency and accuracy, reducing API calls on the HumanEval benchmark by approximately 35\%. Furthermore, the multi-turn iterative debate protocol and reviewer-guided debugging loop significantly enhance the robustness of the generated code.

Experimental results demonstrate that DebateCoder achieves a Pass@1 accuracy of 70.12\% on HumanEval, substantially outperforming the direct generation baseline (59.15\%) and the established MapCoder framework. Ablation studies further validate that the pre-generation debate and post-generation debugging modules function as orthogonal and complementary strategies for maximizing accuracy. In conclusion, DebateCoder proves that with well-designed collaborative protocols, small-parameter models can exhibit exceptional code generation capabilities in resource-constrained environments, offering a viable new pathway for the lightweight deployment of multi-agent systems.

\bibliographystyle{unsrt}  
\bibliography{references}

\end{document}